# Blockchain access control Ecosystem for Big Data security

Uchi Ugobame Uchibeke[*]
*Department of Computer Science*
*University of Saskatchewan*
Saskatoon, Canada
uchi.u@usask.ca

Sara Hosseinzadeh Kassani[*]
*Department of Computer Science*
*University of Saskatchewan*
Saskatoon, Canada
sara.kassani@usask.ca

Ralph Deters
*Department of Computer Science*
*University of Saskatchewan*
Saskatoon, Canada
deters@cs.usask.ca

*Abstract*— In recent years, the advancement in modern technologies has experienced an explosion of huge data sets being captured and recorded in different fields, but also given rise to concerns the security and protection of data storage, transmission, processing, and access to data. The blockchain is a distributed ledger that records transactions in a secure, flexible, verifiable and permanent way. Transactions in a blockchain can be an exchange of an asset, the execution of the terms of a smart contract, or an update to a record. In this paper, we have developed a blockchain access control ecosystem that gives asset owners the sovereign right to effectively manage access control of large data sets and protect against data breaches. The Linux Foundation's Hyperledger Fabric blockchain is used to run the business network while the Hyperledger composer tool is used to implement the smart contracts or transaction processing functions that run on the blockchain network.

## I. Introduction

Due to the rapid development of technology, large amounts of heterogeneous data generated every day which surpassed the capacity to query, processes, and retrieves data effectively. This big data era has opened doors, providing opportunities that were not available before [1]. At the same time, data many applications are becoming diverse in sizes and formats, accordingly poses significant challenges since more flexible data processing tools and platforms are needed [2][3].

Most of the new big data solutions are built on Cloud Computing Technology, which has led to the development of various computation services to accommodate the drastic increase in volume and velocity of big data [4]. The essence of cloud computing environments is to provide significant benefits, such as cost-effective and reliable, scalable and distributed computing of large datasets over the Internet or over a network in both business and academic fields [5]. However, the adoption of distributed clouds is increasingly posing data management security and privacy challenges [6] such as data ownership, multi-tenancy and access control issues since the provided infrastructure is not owned by the users [7].

The Plant Phenotyping and Imaging Research Centre (P$^2$IRC) project at University of Saskatchewan combine the large data sets with different sources and formats from plant genomics, phenotypes, agriculture and crop images for designing crops, and developing innovative methods for global food security. There are a variety of team of researchers and organizations in this project from the plant and agriculture science to bioinformatics and computing science that provide different types of data. With the increasing demand for interactions across these disparate organization with heterogeneous data sources has motivated researchers to provide a scalable cloud-based architecture for storing and processing large number of data source available, and also handling different access control patterns in order to provide secure and reliable answers to the user-specific queries in this project [8].

Blockchain is a cryptographic technology that is used to distribute users' metadata into a secure and distributed database. Each block holds records both for the transactions and data about the block. A secure hash function is used in each node as a reference for validation of data in the existing blocks and if the hash succeeds to be consistent with the data, the block accepted to join to the chain of blocks. All the processing on a blockchain is recorded as traceable transactions carried out between users that serves as a proof of ownership. Once the transaction considered complete it is stored as a permanent record in the blockchain [9] [10].

All nodes in the blockchain network have a full history of the entire blockchain and recently joined nodes must synchronize the full blockchain content in order to ensure trustworthiness and credibility of data. Each node in the blockchain technology keep replicas of the data and agree on an execution order of transactions, therefore, it is almost impossible to attack and fraud against the blockchain network since there is no single point of failure. The primary application for blockchains extended from digital currencies to specific domain applications such as Internet of Things (IoT), supply chain management, networking, healthcare and medical science, cloud computing etc. [11] [12].

The rest of this paper is organized as follows. Section 2 describes the data security and access control issues that need to be solved in order to provide secure data management platforms. Section 3, describes the architecture of proposed system and the security infrastructure. Section 4, discuss the blockchain security solutions that are being used in the area of big data. Finally, in Section 5, we present our conclusions.

---

[*]: These authors contributed equally.

## II. BIG DATA SECURITY AND PRIVACY ISSUES

Cloud computing technology is considered to be the top most strategic technology for enabling organizations to reduce costs and add flexibility to their services in order to handle big data for both computing and storage [13]. However, cloud computing still needed to address security and privacy issues such as data breaches, data exposure, and malicious adversary or even by cloud users [14] [15]. Consequently, cloud providers do not ensure the levels of protection required for appropriate big data security and privacy. This means that the aforementioned concerns related to security and privacy must be taken into consideration in adoption of cloud computing services. Some of the key challenges in supporting large-scale data sets in cloud computing environments are:

### A. Data Privacy Preservation

Privacy-preserving methods should ensure secure access to users' private data without compromising sensitive information from malicious threads. Data privacy is an important necessity in any secured system since customers' data is considered as an asset to both individuals and organizations [16] [17]. Privacy-sensitive information about users in the cloud computing environment is capturing and recording from a wide range of different areas and disciplines. Data concerning medical records or consumer's credit card information are some of the typical examples of the sensitive information. Hence, there is a need to enforce privacy-preserving policies and requirements to protect private information against disclosure using access control mechanisms [18] [19].

### B. Authentication and Authorization

Ensuring authentication and flexible authorization capabilities for providing a secure access control to the data and resources in a multi-user environment with a reliable mechanism is necessary [20]. Shared data is often more vulnerable to unwanted disclosure, security threats, irreversible losses, fraudulent activities, suspicious behavior and attacks [21]. A central cloud security mechanism should have implemented to allow only designated users could have access to data based on their identity, along with roles and access control rules for stored identities.

### C. Identity and Access control Management

Cloud environments are generally based on potentially untrustworthy multi-user environment and, hence, security solutions should be considered for protection of data in such environments [22], for example, ensuring secure authentication, authorization, delegation, data privacy, data confidentiality, and integrity will aid in providing efficient access control to both, over the large-scale data and cloud services [23]. Protecting information about personal identity within the cloud that can be exploited by an attacker aiming to find out the identity of the person is another important concern that needs to be addressed. Identity theft is one of the main challenges that arise from the fraud, misuse of identities, both users' sensitive data and confidential data in the cloud. One aspect that makes detecting identity theft difficult for investigators is the uncertainty on verifying the legitimate user's identity [24].

Generally, storing and accessing big data in the cloud has always been challenging, since it should be secure, highly available and also provide efficient data access control. However, these goals inevitably conflict with each other. Accordingly, we need to improve the data access control performance while providing required security assurance and system availability. In distributed systems, the distribution of data on multiple nodes has a significant impact on data availability, as well as its security. However, highly faulting tolerant and highly availability might not result in high-security assurance or high access efficiency. For example, with more replicas in a system, the system gain fault-tolerant capability leading to improvement of the system reliability and performance of write and read operations, but no result in higher security assurance. Thus, it is essential to design a system that properly meets the necessary requirements in terms of the efficient access control system, availability, and security assurance in the cloud [25].

### D. Data Ownership

Big Data storage and processing services is concerned with the issues of data sharing particularly in cloud environments due to untrustworthy, multi-ownership and dynamic nature of these environments which demands new approaches to architecture, tools, and practices. People and groups often feel that they 'own' organizational data and perceive the data as an asset of their organization, which often results in data ownership problems [14].

Providing proper data ownership roles support integrity and confidentiality of data and privacy of participants from the third party while assuring the accessing data only to users with legitimate access to data. The data ownership issue is exacerbated in cloud environments because of the necessity of accessing data from multiple endpoints which have various ownerships and data access policies, also centralized modeling methods are inefficient to accommodate data ownership issues and establish dynamic relationships of trust [26].

### E. Policy Management

Establishing consistent rules for monitoring and controlling certain actions for protecting the integrity and confidentiality of data stored in a storage system and shared by multiple clients is another issue to be addressed [27]. To increase protection against threats such as disclosure, misuse of data, privacy invasions, it is important for policy makers to design a unified and efficient access control policy management framework that enforces policies for securing both data and its provenance. To achieve this goal, we need an access control policy management system that enforces the appropriate access control policies. Role-Based Access Control (RBAC) models have become the most attractive access control model by simplifying the management of security policies using roles, especially in large enterprise systems. The basic notion of RBAC mechanism is that the access permission to data is restricted based on users assigned roles [28] [29].

With the advent of high-throughput genomics, scientific domains such as life scientists are starting to encounter challenges in handling the size and structure of data. Due to the inefficient traditional SQL model, researchers have been looking for a better approach and NoSQL, with their potential to deal with large, heterogeneous and dynamic schema, are gaining attention as an alternative approach. NoSQL databases, such as Apache Cassandra, MongoDB, DynamoDB address this challenge by handling the vast amounts of semi-structured and unstructured data by providing scalable methods for both storing and answering queries. For the P2IRC project we have used Apache Cassandra to store biological sequence data for multi-omics data and metadata. Fig. 1 is the representation of P2IRC project data architecture and demonstrate how a client send query, data analysis and visualization request to NoSQL database.

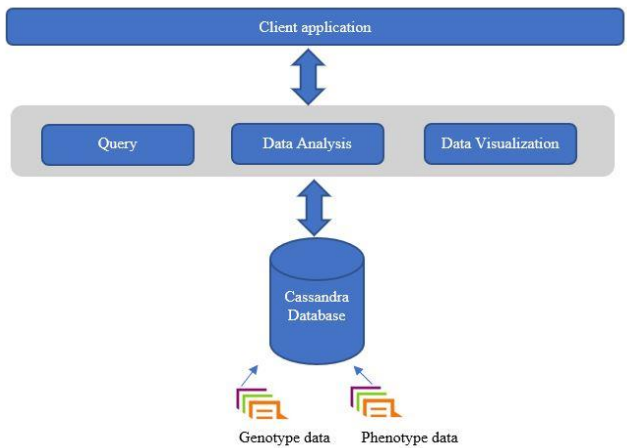

Fig.1: An overview of P2IRC project data architecture

Fig. 2 depict the demonstration of P2IRC data cluster in an Apache cassandra node.

```
Connected to P2IRC Cluster at 127.0.0.1:9042.
[cqlsh 5.0.1 | Cassandra 3.11.0 | CQL spec 3.4.4 | Native protocol v4]
Use HELP for help.
cqlsh>
```

Fig.2: Data cluster in Apache Cassandra for P2IRC project

Fig. 3. illustrates part of CQL code to import genotype data and structure in to Apache Cassandra table.

```
CREATE TABLE P2IRC_genotype(chrom text,pos int,id text,ref text,alt text,qual text,filter
text,info text,format text,s88 text,s108 text,s139 text,s159 text,s265 text,s350 text,s351
text,s403 text,s410 text,s424 text,s428 text,s430 text,s470 text,s476 text,s484 text,s504
text,s506 text,s531 text,s544 text,s546 text,s628 text,s630 text,s680 text,s681 text,s685
text,s687 text,s728 text,s742 text,s763 text,s765 text,s766 text,s768 text,s772 text,s801
text,s853 text,s854 text,s867 text,s868 text,s870 text,s915 text,s932 text,s991 text,s992
text,s997 text,s1002 text,s1006 text,s1061 text,s1062 text,s1063 text,s1066 text,s1070
text,s1158 text,s1166 text,s1254 text,s1257 text,s1313 text,s1317 text,s1552 text,s1612
text,s1622 text,s1651 text,s1652 text,s1676 text,s1684 text,s1739 text,s1741 text,s1756
text,s1757 text,s1793 text,s1797 text,s1819 text,s1820 text,s1829 text,s1834 text,s1835
text,s1851 text,s1852 text,s1853 text,s1872 text,s1890 text,s1925 text,s1942 text,s1943
text,s1954 text,s2016 text,s2017 text,s2031 text,s2053 text,s2057 text,s2081 text,s2091
text,s2106 text,s2108 text,s2141 text,s2159 text,s2166 text,s2171 text,s2191 text,s2202
```

Fig.3: Import genotype data into Apache Cassandra table

In P2IRC project, As the volume of data continues to expand by different collaborators, the ability to securely store and access data becomes increasingly important. Current security approaches do not appear to provide enough coverage since the data become too vast and complex to efficiently store, query, analyze, and share among collaborators. We have developed and implemented a blockchain access control ecosystem, to address this growing need of access control management for this project.

### III. BLOCKCHAIN ACCESS CONTROL ECOSYSTEM

Blockchain technology is revolutionizing the way we store and share data and other digital artifacts. It has the potential to digitally-enable businesses and open new lines of businesses. The Hyperledger Fabric blockchain, led by IBM, is a private and permissioned blockchain that provides the benefits of a blockchain in addition to allowing enterprises to control access to the blockchain network [30]. The Hyperledger architecture working group stated that with Hyperledger, the assumption is that it will be running in an environment of partial trust like in a company or between companies that do not completely trust each other.

In this study, the Hyperledger Fabric blockchain is used to implement access control of big data by borrowing from two existing access control paradigms: 1. Identity-Based Access Control (IBAC), and 2. Role-Based Access Control (RBAC). For each of the access control implementation, five operations are implemented: 1. Request access, 2. Grant access, 3. Revoke access, 4. Verify access, and 5. View asset.

#### A. Blockchain Identity-Based Access Control (BIBAC)

In Identity-Based Access Control, access is granted on a user-by-user basis [31]. If a user needs access to an asset, checks are run to verify that the user's identity has access to the asset. With the Hyperledger Fabric implementation in the current study, when a user requests to access an asset, the owner of the asset grants them access. Futhermore, the owner has the ability to revoke the granted access at any point in time. Additionally, the user who requested the access or another party who is authorized to do so can verify the access rights of the requesting user. Once verified, the user can view or access the asset. All these actions: request, grant, revoke, verify and view, are performed via certificate-signed requests to the blockchain.

The Blockchain Identity-Based Access Control Business Network (BIBAC BN) in this study, modeled using Hyperledger composer, consists of a model resource with definitions of a *person* participant, a *data* asset, and a *request*, *grant*, *revoke*, *verify* and *view* transactions as shown in Fig. 4. When a new participant joins the network, they are issued an identity with a key for submitting transactions to the blockchain. With their key, they can issue one of the transactions implemented in the business network. The implementations of the smart contract are done using the Hyperledger composer client API. The business network historian shows a list of all transactions submitted to the blockchain. With the historian, a user is able to see who has viewed or attempted to view assets that they own. This selective transparency allows for auditability and integrity. Hence, protecting against data breach.

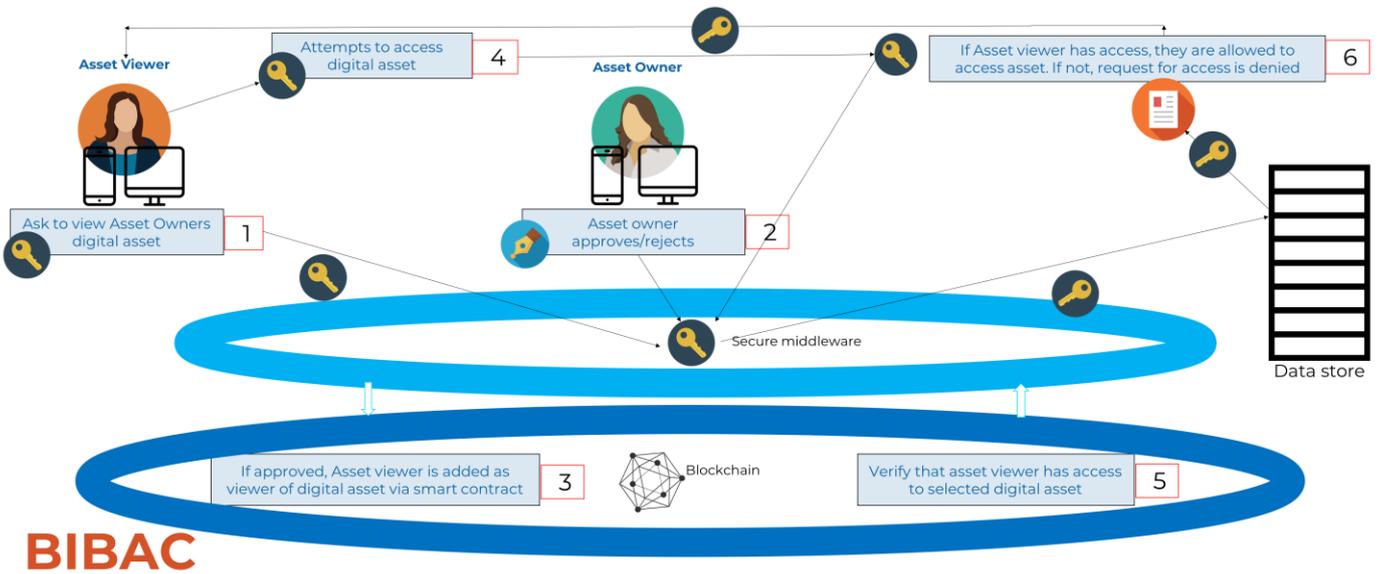

Fig. 4. Model of a Blockchain Identity-Based Access Control (BIBAC) Ecosystem

## B. Blockchain Role-Based Access Control (BRBAC)

In Role-Based Access Control, access is granted on a role-by-role basis. Users are assigned roles and roles are assigned privileges [28] [29]. With the Hyperledger Fabric implementation in this study, a user requests to view an asset and a contract is triggered to pull the roles that have access to that asset. If a user has a role that has access privileges to that asset, that user is allowed to view the asset. Role assignments are controlled by asset owners as shown in Fig. 5.

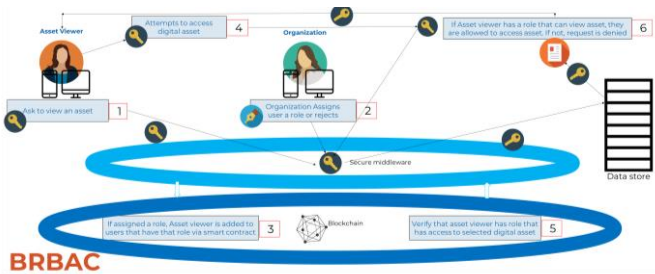

Fig. 5. Role-Based Access Control (BRBAC) Ecosystem

In the Blockchain Role-Based Access Control Business Network (BRBAC BN) in this study, the model resource file contains definitions of participants *person* and *organization*, asset *data* and transactions *request*, *grant*, *revoke*, *verify* and *view* transactions. The assumption is that the data in this Business network is owned by an organization and that persons who work for the organization are granted access privileges to a subset of the big data based on their role in the organization. The organization, using its key or acting through users authorized by the organization, assign roles to users. When a user attempts to access a dataset, the *verify* smart contract or transaction verifies that the user has a role with access rights to that data. Depending on the response from this smart contract, they are allowed or denied access. Fig. 6C is a transaction or smart contract that verifies that a user has access to an asset.

## IV. DISCUSSION

In implementing the prototype of the Blockchain Access Control ecosystem of the current study, we compared two approaches: 1. Control access to data and store the data as part of the Blockchain Business network, and 2. Use the Blockchain Business Network as a vessel to verify that an entity has access to a dataset represented by an ID. The first approach involves developing a big data storage mechanism as part of the Blockchain Business network itself. Additionally, it also means that data has to be migrated from existing systems or that the blockchain network has to be run in parallel with existing data stores that an organization already uses, if a going-forward approach to blockchain adoption is employed. We envision that these limitations will be daunting to organizations because they not only present a cost challenge but a feasibility challenge, also. Considering the above, we decided to go with the second approach: A Blockchain Business Network as a vessel to verify that an entity has access to a dataset represented by an ID. Using a Blockchain Business Network as a vessel to verify that an entity has access to a dataset represented by an ID comes with numerous of benefits. One of the benefits is that there is a low barrier to entry for organizations that need to harness the benefits of blockchain for access control.

```
/**
 * GiveAccess transaction processing function/smart contract.
 * @param {org.assetchain.biznet.GiveAccess} give: Action from user
 * @transaction
 */
function Give(give) {
    var oldValue = give.asset;
    var factory = getFactory();
    // Add new viewers and editors to the asset
    give.asset.viewers = give.viewers.concat(give.viewers);
    give.asset.editors = give.editors.concat(give.editors);
    // Get the asset registry for the asset.
    return getAssetRegistry('org.assetchain.biznet.Resource')
        .then(function (assetRegistry) {
            // Update the asset in the asset registry.
            return assetRegistry.update(give.asset);
        }).then(function () {
        });
}
```

A. *GiveAccess* transaction submitted by asset owner

```
/**
 * CanView transaction processing function/smart contract.
 * @param {org.assetchain.biznet.CanView} CanView Details from user
 * @transaction
 */
function Check(CanView) {
    return getAssetRegistry('org.assetchain.biznet.Resource')
        .then(function (AssetRegistry) {
            return AssetRegistry.get(CanView.asset.assetId);
        }).then (function(asset) {
            if(asset != undefined) {
                return asset.viewers.indexOf(CanView.asset) >= 0
            } else {
                return false
            }
        })
}
```

C. *CanView* transaction submitted when a user attempts to view an asset

```
/**
 * RevokeAccess transaction processing function/smart contract.
 * @param {org.assetchain.biznet.RevokeAccess} revoke: Instructions from user
 * @transaction
 */
function Revoke(revoke) {
    var oldValue = revoke.asset, factory = getFactory();
    revoke.asset.viewers = revoke.asset.viewers.filter(function(viewers){
        return revoke.users.indexOf(viewers) == -1;
    });
    revoke.asset.editors = revoke.asset.editors.filter(function(editors){
        return revoke.users.indexOf(editors) == -1;
    });
    return getAssetRegistry('org.assetchain.biznet.Resource')
        .then(function (assetRegistry) {
            return assetRegistry.update(revoke.asset);
        })
}
```

B. *RevokeAccess* transaction submitted by asset owner to take back access privilege

Fig. 6. Transaction processing function/smart contract code

```
app.post('/giveAccess', function(req, res) {
    const twiml = new MessagingResponse({
        accountSid : config.twilio.accountSid,
        authToken : config.twilio.authToken
    });

    var jsonData = helpers.jsonFromSMS(req.body); // Convert sms to JSON
    var input, sender, sender = req.body.From, msg = jsonData.msg;
    var options = { url : apiBase + "Give", method : 'POST', json : jsonData.input };

    // Make a secure https request to the hyperledger blockchain
    request(options, function(error, response, body) {
        if (!error && response && response.statusCode == 200) {
            console.log("Blockchain Transaction ID: " + response.body.transactionId);
            .
            .
            .
            msg = "RESPONSE SMS TO SEND BACK TO USER"
        } else {
            msg = "ERROR SMS TO SEND BACK TO USER"
        }
        res.writeHead(200, {
            'Content-Type' : 'text/xml'
        });
        twiml.message(msg);
        res.end(twiml.toString()); // Send SMS response to users number
    });
});
```

Fig. 7. Big data access control with Blockchain and SMS

Using either the Blockchain Identity-Based Access Control Business network (BIBAC BN) or the Blockchain Role-Based Access Control Business Network (BRBAC BN) approach described above, organizations can assign IDs to data assets and have the blockchain serve as an auditable access control layer between users and their secure data store. The data IDs can be defined to represent a specific asset, a query that pulls some data or an encoded function that runs to pull data from an existing data store. Additionally, with this flexibility comes modularity, a separation of concerns into different components. The blockchain access management ecosystem is separate from the data store and hence, can be used as an access control layer to existing infrastructure. Additionally, we can enable interaction with the blockchain via SMS as shown in Fig 7. So, a user can text the blockchain to give or revoke access to their asset. Finally, other devices with HTTPS capability can be connected easily to the blockchain.

With these benefits, however, there are a few noteworthy challenges. One of the most prominent challenges is the newness of the Hyperledger Fabric blockchain. The newness of the technology raises questions about stability. Additionally, changes and updates are being made to the fabric distributed ledger and smart contract engine. Over time, however, this challenge will wane. As the technology advances and stability improves, we predict that the benefits of using blockchain for access control will rise and that new opportunities for applying blockchain to big data will emerge.

V. CONCLUSION

In the era of big data, large volumes of heterogeneous data are generated from different sources allowing big data management to play a pivotal role in the success and viability of any businesses. However, the existing solutions for big data access control system is facing several challenges and various threats and risks of data sharing posed by inadequate data security and data privacy models. In this paper, we provide an architecture for access control management by using a decentralized security system based on blockchain. The technology behind blockchain provides a solution to the challenges associated with traditional and centralized access control and ensures data transparency and traceability for secure data sharing.

REFERENCES

[1] P. Cato, P. Gölzer, and W. Demmelhuber, "An investigation into the implementation factors affecting the success of big data systems," in *2015 11th International Conference on Innovations in Information Technology (IIT)*, 2015, pp. 134–139.

[2] D. Xia, Z. Rong, Y. Zhou, B. Wang, Y. Li, and Z. Zhang, "Discovery and analysis of usage data based on hadoop for personalized information access," in *Proceedings - 16th IEEE International Conference on Computational Science and Engineering, CSE 2013*, 2013, pp. 917–924.

[3] V. N. Gudivada, R. Baeza-Yates, and V. V. Raghavan, "Big data: Promises and problems," *Computer*, vol. 48, no. 3, pp. 20–23, 2015.

[4] I. A. T. Hashem, I. Yaqoob, N. B. Anuar, S. Mokhtar, A. Gani, and S. Ullah Khan, "The rise of 'big data' on cloud computing: Review and open research issues," *Inf. Syst.*, vol. 47, pp. 98–115, 2015.

[5] G. Chockler, E. Dekel, J. Jaja, and J. Lin, "Special issue on cloud computing," *Journal of Parallel and Distributed Computing*, vol. 71, no. 6, p. 731, 2011.

[6] B. Varghese and R. Buyya, "Next generation cloud computing: New trends and research directions," *Futur. Gener. Comput. Syst.*, vol. 79, pp. 849–861, 2018.

[7] Y. A. Younis, K. Kifayat, and M. Merabti, "An access control model for cloud computing," *J. Inf. Secur. Appl.*, vol. 19, no. 1, pp. 45–60, 2014.

[8] "Plant Phenotyping and Imaging Research Centre (P2IRC)." .

[9] W. Meng, E. Tischhauser, Q. Wang, Y. Wang, and J. Han, "When Intrusion Detection Meets Blockchain Technology: A Review," *IEEE Access*, 2018.

[10] H. Guo, E. Meamari, and C. C. Shen, "Multi-authority attribute-based access control with smart contract," in *ACM International Conference Proceeding Series*, 2019.

[11] D. D. F. Maesa, A. Marino, and L. Ricci, "Uncovering the bitcoin blockchain: An analysis of the full users graph," in *Proceedings - 3rd IEEE International Conference on Data Science and Advanced Analytics, DSAA 2016*, 2016, pp. 537–546.

[12] H. Guo, E. Meamari, and C.-C. Shen, "Blockchain-inspired Event Recording System for Autonomous Vehicles," Sep. 2018.

[13] M. E. Jeyd and S. M. Hashemi, "Provide a framework for selecting services based on cloud computing technology in e-Capital Market," in *8th International Conference on e-Commerce in Developing Countries: With Focus on e-Trust*, 2014, pp. 1–6.

[14] S. Sahmim and H. Gharsellaoui, "Privacy and Security in Internet-based Computing: Cloud Computing, Internet of Things, Cloud of Things: A review," in *Procedia Computer Science*, 2017, vol. 112, pp. 1516–1522.

[15] M. B. Mollah, M. A. K. Azad, and A. Vasilakos, "Security and privacy challenges in mobile cloud computing: Survey and way ahead," *Journal of Network and Computer Applications*, vol. 84. pp. 38–54, 2017.

[16] R. Akalu, "Privacy, consent and vehicular ad hoc networks (VANETs)," *Comput. Law Secur. Rev.*, vol. 34, no. 1, pp. 175–179, 2018.

[17] R. Manjula and R. Datta, "A novel source location privacy preservation technique to achieve enhanced privacy and network lifetime in WSNs," *Pervasive Mob. Comput.*, vol. 44, pp. 58–73, 2018.

[18] A. Sahi, D. Lai, and Y. Li, "Security and privacy preserving approaches in the eHealth clouds with disaster recovery plan," *Comput. Biol. Med.*, vol. 78, pp. 1–8, 2016.

[19] D. Teneyuca, "Internet cloud security: The illusion of inclusion," *Inf. Secur. Tech. Rep.*, vol. 16, no. 3–4, pp. 102–107, 2011.

[20] R. Li, H. Asaeda, J. Li, and X. Fu, "A distributed authentication and authorization scheme for in-network big data sharing," *Digit. Commun. Networks*, vol. 3, no. 4, pp. 226–235, 2017.

[21] K. Selvamani and S. Jayanthi, "A review on cloud data security and its mitigation techniques," in *Procedia Computer Science*, 2015, vol. 48, no. C, pp. 347–352.

[22] I. Weber, S. Nepal, and L. Zhu, "Developing Dependable and Secure Cloud Applications," *IEEE Internet Comput.*, vol. 20, no. 3, pp. 74–79, 2016.

[23] Q. G. K. Safi, S. Luo, C. Wei, L. Pan, and G. Yan, "Cloud-based security and privacy-aware information dissemination over ubiquitous VANETs," *Comput. Stand. Interfaces*, vol. 56, pp. 107–115, 2018.

[24] A. Singh and K. Chatterjee, "Cloud security issues and challenges: A survey," *J. Netw. Comput. Appl.*, vol. 79, pp. 88–115, 2017.

[25] E. Koutrouli and A. Tsalgatidou, "Taxonomy of attacks and defense mechanisms in P2P reputation systems-Lessons for reputation system designers," *Computer Science Review*, vol. 6, no. 2–3. pp. 47–70, 2012.

[26] J. Li, X. Chen, Q. Huang, and D. S. Wong, "Digital provenance: Enabling secure data forensics in cloud computing," *Futur. Gener. Comput. Syst.*, vol. 37, pp. 259–266, 2014.

[27] F. H. Alqahtani, "Developing an Information Security Policy: A Case Study Approach," in *Procedia Computer Science*, 2017, vol. 124, pp. 691–697.

[28] J. D. Ultra and S. Pancho-Festin, "A simple model of separation of duty for access control models," *Comput. Secur.*, vol. 68, pp. 69–80, 2017.

[29] F. Rezaeibagha and Y. Mu, "Distributed clinical data sharing via dynamic access-control policy transformation," *Int. J. Med. Inform.*, vol. 89, pp. 25–31, 2016.

[30] H. A. Working and G. (WG), "Hyperledger Architecture," vol. 1, 2017.

[31] A. R. Khan, "Access control in cloud computing environment," *ARPN J. Eng. Appl. Sci.*, vol. 7, no. 5, pp. 613–615, 2012.